\documentclass[useAMS,usenatbib]{mn2e}
\usepackage{graphicx,amssymb}
\usepackage[T1]{fontenc}
\usepackage{aecompl}

\def\eb{$\epsilon_{\rm B}$}
\def\ee{$\epsilon_{\rm e}$}
\def\csi{$\xi_{\rm e}$}
\def\etag{$\eta_\gamma$}

\def\eiso{$E_{\rm prompt}$}
\def\llat{$L_{LAT}$}
\def\lovere{$L_{LAT}/E_{\rm prompt}$}

\title[Clustering of LAT light curves]
  {Clustering of LAT light curves: a clue to the origin of high-energy emission in Gamma-Ray Bursts}
\author[L. Nava et al.]
  {L.~Nava$^{1,2}$\thanks{lara.nava@mail.huji.ac.il},
  G.~Vianello$^3$,
  N.~Omodei$^3$,
  G.~Ghisellini$^4$,
  G.~Ghirlanda$^4$,
  A.~Celotti$^{5,4,6}$,\newauthor  
  F.~Longo$^{6,7}$,
  R.~Desiante$^{7,8}$,
  R.~Barniol Duran$^{1}$
\\
 \\    
$^1$Racah Institute of Physics, The Hebrew University of Jerusalem, 91904, Israel\\
$^2$APC, Univ Paris Diderot, CNRS/IN2P3, CEA/Irfu, Obs de Paris, Sorbonne Paris Cit\'e, France\\
$^3$W. W. Hansen Experimental Physics Laboratory, Kavli Institute for Particle Astrophysics and Cosmology, Department of Physics \\~~and SLAC National Accelerator Laboratory, Stanford University, Stanford, CA 94305, USA\\
$^4$INAF--Osservatorio Astronomico di Brera, via E. Bianchi 46, I-23807 Merate, Italy\\
$^5$SISSA, via Bonomea 265, I-34136 Trieste, Italy\\
$^6$INFN, Sezione di Trieste, I-34127 Trieste, Italy\\
$^7$Dipartimento di Fisica, Universit\`a di Trieste, I-34127 Trieste, Italy\\
$^8$Universit\`a di Udine, Via delle Scienze 208, I-33100 Udine, Italy
}
\date{Released}

\pagerange{\pageref{firstpage}--\pageref{lastpage}} \pubyear{2012}

\begin{document}
\voffset -1truecm 
\label{firstpage}

\maketitle

\begin{abstract}
The physical origin of the $>0.1\,$GeV emission detected from Gamma-Ray Bursts (GRBs) by the {\it Fermi} satellite has not yet been completely understood.
In this work we consider the GeV light curves of ten GRBs with measured redshift detected by the \hbox{{\it Fermi-}LAT}. These light curves are characterised by a long-lived ($\gtrsim10^2$ seconds) emission, whose luminosity decays in time as a power-law. While the decay rate is similar for all GRBs (i.e. $L_{\rm LAT}\propto t^{-1.2}$), the normalisation spans about two orders of magnitude in luminosity. However, after re-normalising the luminosities to the prompt energetics \eiso\ the light curves overlap.
We consider the scenario in which the temporally extended LAT emission is dominated by synchrotron radiation from electrons accelerated at the forward external shock. According to this model, at high-energies (i.e. above the typical synchrotron frequencies) a small dispersion of the \eiso-normalised light curves is expected. The fact that the LAT temporally extended emission follows this behaviour reinforces its interpretation in terms of afterglow radiation from external shocks. Assuming this scenario, we argue that the parameters \ee\ and \etag\ (i.e., the fraction of shock-dissipated energy gained by the electrons, and the efficiency of the mechanism producing the prompt radiation, respectively) must be narrowly distributed.

\end{abstract}


\section{Introduction}\label{introduction}
Since the beginning of observations in August 2008, the {\it Fermi} Large Area Telescope (LAT; \citealt{atwood09}) has observed significant emission above 0.1 GeV from about 60 GRBs \footnote{http://fermi.gsfc.nasa.gov/ssc/observations/types/grbs/lat\_grbs\\/table.php}. 
The redshift has been measured for 16 out of 60 GRBs, and it ranges from $z$=0.145 (GRB 130702A) to $z$=4.35 (GRB 080916C). 
Except for very faint events, the emission detected by the LAT above 0.1 GeV lasts hundreds to thousands of seconds, much longer than the prompt emission detected by the {\it Fermi} Gamma-Ray Burst Monitor (GBM; \citealt{meegan09}). 
{\it Fermi}-LAT revealed that the flux of this temporally extended emission decays in time as a power-law $t^{-\alpha}$, with temporal index around $\alpha=1.2$ \citep[ACK13 hereafter]{latcatalog}. The spectral analysis of the LAT data alone showed that spectra can be modelled with a power-law function $dN/dE\propto E^{-\Gamma}$ with photon index $\Gamma$ between 2 and 2.1 \citep[][ACK13]{ghisellini10}.
In six cases, the spectral modelling of the GBM and LAT data during the prompt emission phase revealed that an extra-component in the spectrum, apart from the canonical Band function, must be introduced to properly describe the LAT data (ACK13).

The nature of this emission is still not completely understood. The most promising models interpret this emission as radiation from electrons accelerated at the external shock. In particular, several authors invoked a synchrotron origin from the forward shock \citep{kumar09,kumar10,gao09,ghisellini10,ghirlanda10,depasquale10}. Attempts to simultaneously model LAT radiation, optical and X-ray data for few bright LAT bursts resulted in a successful modelling (\citealt{kumar10,lemoine13}, but see \citealt{maxham11}). 
However, problems with this interpretation have also been pointed out.
A handful of photons with energies from $10\,$GeV to $\sim100\,$GeV has been detected in some cases. The detection of  photons with such high energy challenges the synchrotron model, since it has been argued that they cannot be produced by the synchrotron mechanism \citep{piran10,130427LAT}.
Some authors proposed that, while the bulk of the emission is produced via the synchrotron mechanism, these few photons may have a different origin, and may be produced via inverse Compton (IC) scattering \citep{wang13}. 
Other authors have proposed the IC mechanism as an explanation for the entire emission detected by LAT, and not only for the few high-energy ($\gtrsim10\,$GeV) photons. In this case, seed photons for IC scattering can be provided by the prompt radiation \citep{beloborodov13} or eventually, at later time, by synchrotron X-ray/optical afterglow radiation \citep{vurm14}.

\cite{ghisellini10} considered the LAT light curves of the four brightest bursts with measured redshift and found that they follow an interesting behaviour: these light curves overlap when the luminosity of the LAT emission is re-normalised to the total isotropic prompt emission energy \eiso. They argued that this behaviour is predicted by the synchrotron/external-shock model and supports the interpretation of the high-energy emission in terms of afterglow radiation. Similar results, in fact, have been derived from the analysis of the X-ray and optical afterglow light curves, and have been used to argue that, in order to explain the tight relation between afterglow luminosity and prompt energetics, a standard value for the efficiencies \ee\ and \etag\ must be invoked \citep{kumar00,kaneko07,berger07,berger13}, where \ee\ is the ratio between the energy of the non-thermal population of the accelerated electrons and the energy dissipated at the forward external shock, while \etag\ is the efficiency in producing the prompt radiation.

In the present paper we test the solidity of the result found by \cite{ghisellini10} by means of a larger sample (10 events) that includes all GRBs with measured redshift and temporally extended emission above 0.1 GeV. The sample is presented in Section \ref{sect:thesample}. We find that the result by \cite{ghisellini10} is confirmed: the dispersion of the light curves of different bursts decreases when the LAT luminosity is re-normalised using \eiso\ (Section \ref{sect:results}). 
In Section \ref{sect:interpretation} we interpret this result in the context of synchrotron afterglow radiation. In this scenario it is possible to use the width of the \lovere\ distribution to constrain the width of the distribution of two parameters entering the afterglow luminosity: the efficiency \etag\ of the prompt and the shock parameter \ee. We discuss in more detail the results inferred on \ee\ and \etag\ in Section \ref{sect:discussion}, and summarise the conclusions of this work in Section \ref{sect:conclusions}.

\section{The sample}\label{sect:thesample}
We select all GRBs with measured redshift for which a temporally extended emission at energies larger than $0.1\,$GeV has been detected by LAT. Ten bursts satisfy these criteria. Nine of them are included in the First {\it Fermi}-LAT GRB catalog (ACK13), while for GRB 130427A the temporal and spectral analysis is reported in \cite{130427LAT}.
For all of them the emission detected above $0.1\,$GeV is temporally extended, i.e. lasts longer than the duration of the prompt emission, as measured by the $T_{90}$ obtained using GBM data. 
To derive the light curves of the high-energy emission we have used the analysis described in ACK13 applied to the ``Pass 7'' {\it Transient} event class of {\it Fermi}-LAT data\footnote{http://fermi.gsfc.nasa.gov/ssc/data/analysis/documentation/\\Pass7\_usage.html}. 
In particular, we have obtained from the authors of ACK13 the light curves for all LAT-detected GRBs with a temporally extended emission.
Using the redshift measurements reported there, we have then computed the rest-frame light curves (left panel of Fig.\ref{fig:clustering}). Luminosities $L_{LAT}$ are provided in the $0.1-10\,$GeV rest frame energy range. The conversion factor between the observed flux reported in ACK13 in the energy range $0.1-10\,$GeV (observer frame) and the $0.1-10\,$GeV energy range (rest frame) has been computed separately for each time interval in each light curve, by assuming a power law spectrum with the appropriate photon index. 
Temporal breaks in the decay of the light curve have been firmly detected in four bright GRBs: 090510, 090902B, 090926A, and 130427A \citep[ACK13]{130427LAT}. In these cases the initial flux decays faster than the post-break flux and is likely dominated by the contribution from the prompt (ACK13). For these GRBs we exclude from our analysis the part of the light curve before the break. Also, we exclude the initial part of the light curve when it is characterised by a rising flux and/or flux variability, since we are interested in investigating the part of the light curve that decays as a power-law. The complete light curves (in the $0.1-10\,$GeV observer frame) can be found in ACK13.

The prompt energetics \eiso\ have been estimated in the $1\,\rm{keV}-10\,\rm{MeV}$ rest frame energy range from the fluences reported in Table 11 in ACK13. Here we are interested in the energetics of the prompt emission only, therefore, for the cases where an extra-component has been observed, we have ignored its contribution to the $1\,\rm{keV}-10\,\rm{MeV}$ GBM fluence, and considered only the contribution from the low-energy component (typically a Band or Comptonized spectrum), reported in the last column of the ``main component'' section in Table 11 of ACK13. For GRB 130427A, we have used the spectral parameters reported in table S1 in the supplementary material of \cite{130427LAT}, excluding the contribution from the extra power-law component.

\section{Results}\label{sect:results}
\begin{figure*}
\includegraphics[scale=0.72]{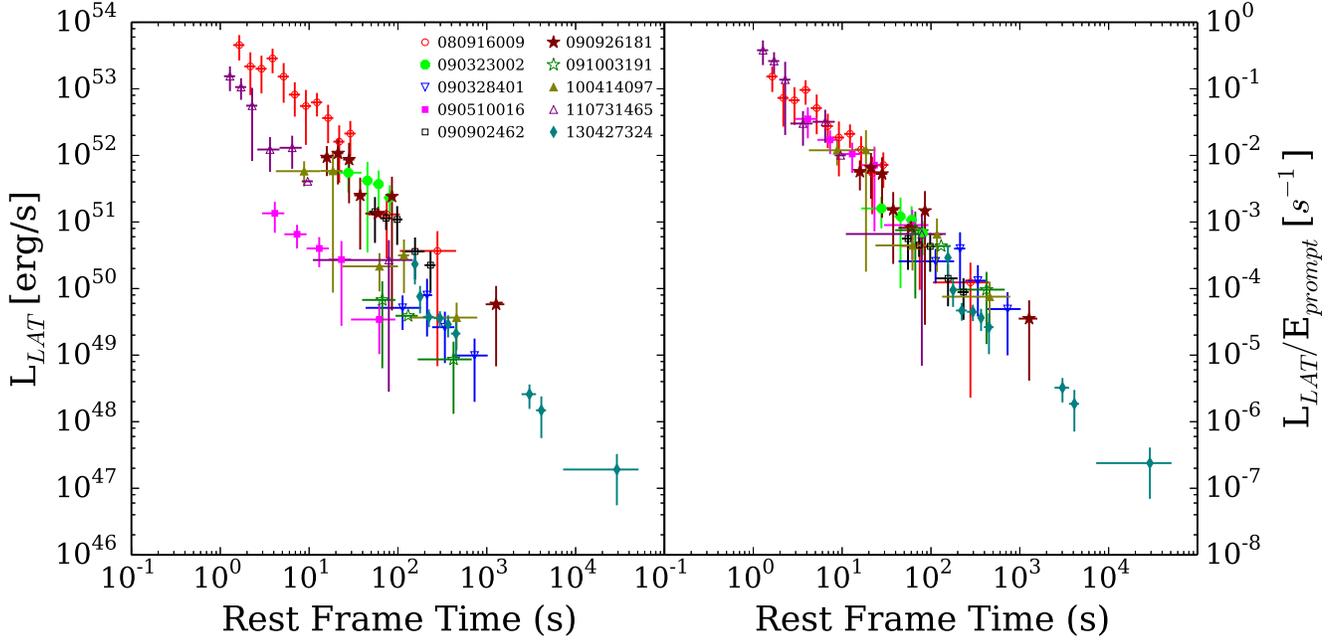}
\caption{Left panel: luminosity ($0.1-10\,\rm{GeV}$, rest frame) as a function of $t$ (the time since the burst trigger, in the rest frame of the central engine) for the ten bursts in our sample. 
Only data belonging to the temporally extended emission are shown.
Right panel: same as in the left panel but the luminosity has been normalised to the prompt energetics $E_{\rm prompt}$, estimated from the GBM in the $1\,\rm{keV}-10\,\rm{MeV}$ energy range.}
\label{fig:clustering}
\end{figure*}
The light curves of all GRBs in our sample share a similar behaviour. 
After an initial phase characterised by a rising flux and/or flux variability (that we excluded from our analysis), 
$L_{LAT}$ decays as a power-law in time: 
$L_{LAT}=K t^{-\alpha}$, where $t$ is the rest frame time since the trigger. 
We refer to this power-law phase as LAT temporally extended emission. 
The light curves of the extended emission are shown in Fig.~\ref{fig:clustering} (left panel).
While the decay rate $\alpha$ is similar among different bursts ($\alpha\sim1.2$, ACK13), the normalisation $K$ spans around two orders of magnitude. 

In the right panel of Fig.~\ref{fig:clustering}, the luminosity of each burst has been divided by \eiso.
For all the events in our sample, the light curves of the extended emissions overlap when they are normalised to the prompt energetics.
The normalisation $K'$ of the different \eiso-normalised light curves (defined by $L_{LAT}/E_{\rm prompt}=K't^{-\alpha}$) is very similar for different bursts and its value spans less than one order of magnitude.
This means that at each given rest frame time $t$ the ratio between the LAT luminosity $L_{LAT}(t)$ and the prompt energetics \eiso\ is roughly the same for all GRBs. 

\begin{figure}
\includegraphics[scale=0.58]{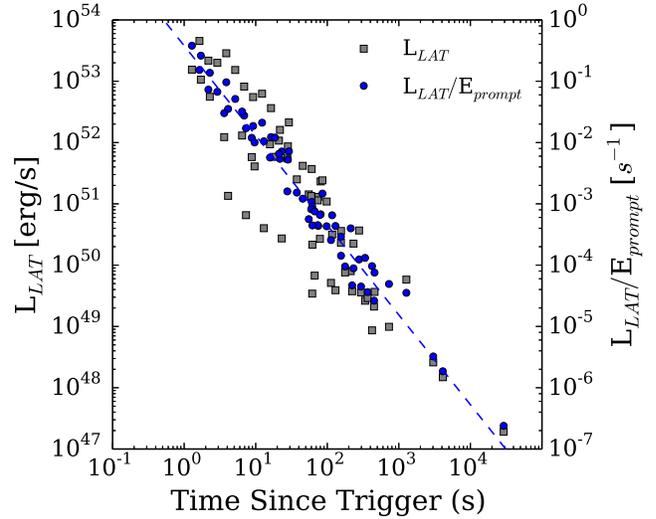}
\caption{Square symbols refer to the LAT luminosity ($y$-axis on the left), while circles refer to the luminosity normalised to \eiso\ ($y$-axis on the right). Only data points belonging to the extended emission phase are shown, without distinguish between different bursts. The dashed line is the best fit of \lovere\ vs $t$, where $t$ is the time since the burst trigger, measured in the rest frame.}
\label{fig:clustering_all}
\end{figure}

In order to quantify the dispersion of the ratio \lovere, in Fig. \ref{fig:clustering_all} we report all data points, 
without distinguishing between different bursts. 
Square symbols refer to LAT luminosities (values are given on the left $y$-axis), while circles refer to LAT luminosities divided by \eiso\ (right $y$-axis). 
In the latter case, data points are less dispersed.  
The vertical dispersion of the blue circles in Fig. \ref{fig:clustering_all} is representative of the average dispersion of the ratio \lovere.
We modelled the distribution of the vertical distances of data points from the best fitting line with a gaussian function. 
We quantify the dispersion of \lovere\ as the standard deviation of this gaussian distribution and find $\sigma_{log(L/E)}=0.23$. 
The best fitting line is shown in Fig. \ref{fig:clustering_all} as a dashed line.

Before proceeding with the analysis and interpretation of this result, we recall that this behaviour (that we refer to as clustering) is not found when \llat\ is normalised to other quantities. 
Intuitively, a clustering is expected if \llat\ is divided by $E_{LAT}= \int L_{LAT} dt$, the total energy emitted in the LAT energy range integrated over the whole duration of the extended emission. If all light curves start more or less at the same time and decay at the same rate, the luminosity at some time $t$ is proportional to the total energy output $E_{LAT}$ and the proportionality constant is the same for all bursts.
Following this reasoning, the clustering of \lovere\ could be explained as the result of two effects: the obvious clustering of $L_{LAT}/E_{LAT}$ and the existence of a (strong) correlation between $E_{LAT}$ and \eiso.
This possibility has been investigated in \cite{nava13}, but only a modest decrease in the dispersion has been found when $L_{LAT}$ is normalised to $E_{LAT}$ and it cannot be the cause of the much stronger clustering found when \eiso\ is used in place of $E_{LAT}$ (see figure 1 in \citealt{nava13} for details, and the text for the discussion).
We also tested if a clustering can be obtained by normalising the LAT luminosity to the peak luminosity and/or to the spectral peak energy of the prompt emission. In the first case the dispersion is slightly reduced \citep{nava13}, while in the second case it remains unaltered.

\section{Forward shock emission from external shocks}\label{sect:interpretation}
In this section we show that the overall properties of the LAT emission and in particular the clustering of the \eiso-normalised light curves are consistent with synchrotron radiation from the forward shock driven by a relativistic blast-wave into the external medium.
To describe the synchrotron emission from forward shock we follow the prescriptions given in \cite{granot02}.
First, we consider the case of an adiabatic blast-wave decelerating into a medium with constant number density $n$. 
At the end of this section we discuss the case of a medium with density $n\propto r^{-2}$ and we show that our results and conclusions are independent from the radial profile of the circum-burst medium.
We assume that the Compton $Y$ parameter is small, so that $(1+Y)\sim1$. In the next section we will demonstrate that this is a good approximation for electrons emitting in the LAT energy range, and that cooling via SSC does not affect the results derived in this section.

During the deceleration phase, the rest frame cooling energy $h\nu_c$ and the injection energy $h\nu_m$ of the synchrotron spectrum are given by:
\begin{equation}\label{eq:Ec}
h\nu_{\rm c}\simeq7.1\times10^{-7} \epsilon_{\rm B,-2}^{-3/2}\, E_{\rm K,54}^{-1/2}\, n^{-1} t^{-1/2}\, {\rm GeV}
\end{equation}
\begin{equation}\label{eq:Em}
h\nu_{\rm m}\simeq0.017\left[\frac{f(p)}{f(2.2)}\right]^2\epsilon_{\rm B,-2}^{1/2} \epsilon_{\rm e,-1}^2\,\xi_{\rm e,-1}^{-2} E_{\rm K,54}^{1/2}\, t^{-3/2}\, {\rm GeV}
\end{equation}
where $t$ is the rest frame time in seconds, $E_{\rm K}$ is the energy content of the fireball, $n=const$, and the notation $Q=10^xQ_x$ is adopted. 
For the microphysical parameters describing the physics of the shock the fiducial parameters commonly adopted since the first papers published on broad-band modelling of afterglow radiation \citep{panaitescu00,panaitescu01,wijers99} are $\epsilon_B=10^{-2}$ (although with a large dispersion) and $\epsilon_e=10^{-1}$. We choose to normalise \eb\ and \ee\ to these reference values.
The normalisation of $h\nu_m$ depends on $f(p)\equiv(p-2)/(p-1)$, where $p$ is the power-law index of the Lorentz factor distribution of the accelerated electrons: $dN_e/d\gamma\propto\gamma^{-p}$. 
The parameter \csi\ (the fraction of electrons that are accelerated into a power-law energy spectrum) is introduced to account for the possibility that not all electrons are efficiently accelerated to a non-thermal energy distribution. 

According to these estimates, even at very early time the energy range of interest (0.1-10 GeV) lies most likely in the high-energy part of the synchrotron spectrum, above $h\nu_c$ and $h\nu_m$.
In both the fast cooling ($\nu_c<\nu_m$) and slow cooling regime ($\nu_c>\nu_m$), the specific luminosity $L_{\rm E}$ for $E > max(h\nu_{\rm c},h\nu_{\rm m})$  is given by:
\begin{eqnarray}\label{eq:mocro_lum}
L_{\rm E}\simeq6.1\times 10^{51}\left[\frac{f(p)}{f(2.2)}\right]^{p-1}\left[\frac{E}{1\,\rm GeV}\right]^{-\frac{p}{2}} \epsilon_{\rm B,-2}^{\frac{p-2}{4}}\epsilon_{\rm e,-1}^{p-1} \xi_{\rm e,-1}^{2-p} \,\nonumber \\ 
\times E_{\rm K,54}^{\frac{p+2}{4}} t^{-\frac{3p-2}{4}} {\rm erg/s/GeV}
\end{eqnarray}
where $E$ is the rest frame photon energy. 
Since on average the observed light curves decay in time as $t^{-1.2}$, observations suggest $p\simeq2.2$, which in turn implies a spectral index $p/2\simeq1.1$ ($L_{\rm E}\propto E^{-p/2}$), in good agreement with the typical spectral indices derived from the spectral analysis of LAT data (ACK13).
Assuming $p=2.2$, the luminosity in the $0.1-10\,$GeV (rest frame) energy range is:
\small
\begin{equation}\label{eq:integrated_lum}
L_{\rm LAT,52}\simeq2.8\left[\frac{f(p)}{f(2.2)}\right]^{1.2}\epsilon_{\rm B,-2}^{0.05} \epsilon_{\rm e,-1}^{1.2} \xi_{\rm e,-1}^{-0.2} E_{\rm K,54}^{1.05}\, t^{-1.15} {\rm erg/s}
\end{equation}
\normalsize

Equation \ref{eq:integrated_lum} shows that the standard afterglow model predicts that the synchrotron luminosity emitted in the LAT energy range is proportional to the energy content of the fireball $E_{\rm K}$, it has a very weak dependence on \eb\ and \csi, and it does not depend on $n$ \citep{kumar00}.
The energy $E_{\rm K}$ is related to the energy emitted in $\gamma$-rays and to \etag\ (the overall efficiency of the mechanism producing the prompt radiation) by the following equation:
\begin{equation}\label{eq:eta}
E_{\rm K}=E_{\rm prompt}\left(\frac{1-\eta_\gamma}{\eta_\gamma}\right)
\end{equation}
In the previous estimates, we adopted $E_{\rm K}=10^{54}E_{K,54}$ since the average value of \eiso\ for our sample is a few $\times10^{53}\,$erg, which for \etag\ ranging between 10-30 per cent gives a kinetic energy $E_{\rm K}\sim10^{54}\,$erg.

\begin{figure}
\includegraphics[scale=0.47]{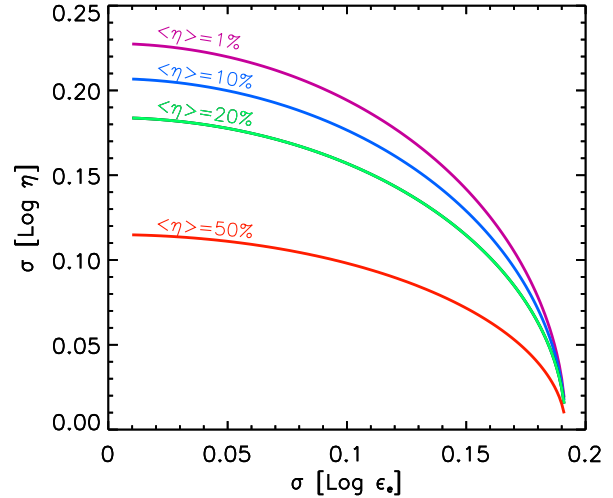}
\caption{Standard deviation for the distributions of $Log\,$\etag\ and $Log\,$\ee\ for different values of the average $\langle\eta_\gamma\rangle$. The two parameters are assumed to be lognormally distributed and uncorrelated. For each value of $\sigma_{Log\,\epsilon}$, the value of $\sigma_{Log\,\eta}$ is inferred from the requirement that the standard deviation of the distribution of  $Log\,$[$\epsilon_{\rm e}^{1.2}$(1-\etag)/\etag] is $\sigma=0.23$.}
\label{fig:dispersion}
\end{figure}

By replacing Eq.~\ref{eq:eta} into Eq.~\ref{eq:integrated_lum} and neglecting non-relevant terms and weak dependencies, we see that in the standard external shock model the ratio between the LAT luminosity and the prompt energetics is mainly a function of \ee\ and \etag:
\begin{equation}\label{eq:ratio}
\frac{L_{\rm LAT}}{E_{\rm prompt}}\propto \epsilon_{\rm e}^{1.2}\, \frac{1-\eta_\gamma}{\eta_\gamma}\, t^{-1.2}
\end{equation}
In this scenario, the dispersion of the ratio \lovere\ is caused by the width of the distributions of \etag\ and \ee. Also other parameters (in particular $p$ and \csi) may give a non-negligible contribution to the dispersion. However, in order to perform a conservative analysis, we assume that all the dispersion has to be ascribed to \etag\ and \ee.
If we assume that these two parameters are independent variables, then the clustering found in the LAT data implies that both \ee\ and \etag\ must be narrowly distributed around a typical value. 
From the data we inferred $\sigma_{Log L/E}=0.23$ (see Section~\ref{sect:results}).
Since it is not possible to disentangle between the contribution of \ee\ and \etag\ to the total dispersion of \lovere, we derive the width of one parameter as a function of the width of the other one, under the assumption that they both have a lognormal distribution and they are uncorrelated (Fig.~\ref{fig:dispersion}). When the contribution of one parameter is assumed to be negligible (i.e. $\sigma\sim0$), the plot shows the maximum width of the distribution of the other parameter. 
For \ee\ the maximum width is simply given by $\sigma=\sigma_{Log L/E}/1.2\sim 0.19$, while for \etag\ the maximum width depends on the average value: it is $\sigma\sim 0.23$ for very small \etag, since in this case $(1-\eta_\gamma)/\eta_\gamma\sim1/\eta_\gamma$, and it is even smaller if the mean value is larger.

A narrow distribution implies that these parameters assume similar values for the GRBs in our sample.
To derive these typical values we compare Eq.~\ref{eq:integrated_lum} with the best fit of the data points (solid line in Fig.~\ref{fig:clustering_all}).
Again, from this analysis it is not possible to separately infer the mean value of each parameter, but we note that they are consistent with the typical values commonly adopted: \ee=0.1 and \etag=0.2.
However, we warn that any attempt to derive the typical values of \ee\ and \etag\ is model dependent.
The normalisation factor in Eq.~\ref{eq:integrated_lum} may change depending on the model adopted to describe the synchrotron emission. Several models are available in the literature  \citep{sari98,panaitescu00,granot02,wijers99}, all giving the same results in terms of dependence of the afterglow synchrotron luminosity on the model parameters, but with a normalisation that can differ even by a factor of $\sim$10, depending on the adopted descriptions (see e.g. \citealt{granot02} for a discussion about the origin of these discrepancies).
Moreover, a different choice of $p$ can introduce a factor of $\gtrsim2$ of difference (for $p$ ranging from 2.2 to 2.5), while the dependence on \csi\ and \eb\ is weaker and can be neglected.  
The estimate of the maximum dispersions of \ee\ and \etag\ is instead model independent and quite robust. If also $p$, \csi\ and \eb\ contribute to $\sigma_{Log L/E}$ (and if all these quantities are uncorrelated), then the inferred width of \ee\ and \etag\ would be even smaller.

\subsubsection*{Wind-like density profile}
We consider an adiabatic blast-wave decelerating in a stratified medium with number density $n=Ar^{-2}$, with $A=3\times10^{35}A_\star\,$cm$^{-1}$. 
In this case, the expression for $h\nu_m$ is the same as the one derived for $n=const$ (Eq.~\ref{eq:Em}), only its normalisation is different by $\lesssim10$ per cent.
The expression for $h\nu_c$ instead is different and, unlike the homogeneous case, $h\nu_c$ increases with time:
\begin{equation}
h\nu_{\rm c}=1.3\times10^{-11} \epsilon_{\rm B,-2}^{-3/2}\, E_{\rm K,54}^{1/2}\, A_\star^{-2} t^{1/2}\, {\rm GeV}
\end{equation}
For typical values of the parameters, it is very unlikely that the cooling frequency could cross the LAT energy range during the temporal window of interest for the LAT emission (i.e., $t\lesssim10^3\rm\,s$), unless the blast-wave is decelerating in a very low-density ambient medium with $A_\star\lesssim6\times10^{-5}\epsilon_{\rm B,-2}^{-3/4}E_{\rm K,54}^{1/4}$.
Therefore, we can safely assume that the LAT energy range always lies above $\nu_c$ and $\nu_m$. 
The equation for the afterglow luminosity at $\nu>max(\nu_c,\nu_m)$ differs from the one derived in the homogeneous medium (Eq.~\ref{eq:integrated_lum}) only by a multiplicative factor of order unity (see e.g. \citealt{panaitescu00}). 
The same conclusions derived in the case of an homogeneous medium are also valid for a wind-like density environment. 
The fact that the afterglow luminosity in the high-energy part of the synchrotron spectrum is insensitive to the value of the density and to its radial profile implies that the empirical clustering found in the data does not help us to discriminate among homogeneous and stratified circum-burst media.
In both cases the theory predicts approximately the same value for the ratio \lovere\ and the very same dependence on the unknown parameters. 
This can explain why the short GRB included in our sample (GRB 090510) does not show any peculiar behaviour: even if its luminosity is low as compared to the average luminosity of the other (long) GRBs included in our sample (as commonly observed for the afterglow luminosity of a short GRB), its \eiso-normalised light curve is perfectly consistent with the \eiso-normalised light curves of long bursts.

\subsection{Synchrotron Self Compton}\label{sect:SSC}
The equations derived in the previous section are based on the assumption that energy losses via SSC cooling are negligible. This is true when $\epsilon_e\lesssim\epsilon_B$. 
However, the parameter \eb\ is perhaps one of the most uncertain parameters of the external shock physics. 
The value \eb=0.01 is typically used as the fiducial value, but the modelling of the afterglow data showed that \eb\ varies over many orders of magnitude, ranging from $10^{-5}$ to $10^{-1}$, and recent studies suggest that it can assume even smaller values \citep{barniolduran13,santana14}.
If $\epsilon_{\rm B}\ll\epsilon_{\rm e}$ electron cooling via inverse Compton scattering of synchrotron photons might be important, especially at early time, during the fast cooling stage \citep{sari01}.
SSC can invalidate our previous results if: 
i) the SSC cooling modifies the overall shape of the synchrotron spectrum and suppresses the synchrotron flux at the relevant energies 0.1-10 GeV, and/or 
ii) if SSC radiation contributes (or even dominates) the emission in the LAT energy range.
However, the relevance of the SSC effects can be attenuated, especially for high-energy photons, by the Klein-Nishina (KN) limit.
Below we estimate the effects of the SSC mechanism on the results derived in the previous section.
A proper estimate of the KN limit for the energy range of interest will be considered. 
Our description of the SSC mechanism is mainly based on the work by \cite{nakar09} and \cite{wang10}.

First, we consider the Thomson scattering regime. In this case the Compton $Y$ parameter is constant (i.e., it assumes the same value for all the emitting electrons) and the cooling frequency is reduced by a factor $(1+Y)^2$: $\nu_{\rm c}=\nu_{\rm c}^{syn}/(1+Y)^2\simeq\nu_{\rm c}^{syn}(\epsilon_{\rm B}/\epsilon_{\rm e})$, where the last expression is valid in fast cooling and $Y\gg1$, and $\nu_{\rm c}^{syn}$ is the cooling frequency when the SSC is not important and its expression is given by Eq.~\ref{eq:Ec}.
The flux at $\nu>max(\nu_c,\nu_m)$ is also reduced, by a factor $(1+Y)\simeq Y\simeq\sqrt{\epsilon_{\rm e}/\epsilon_{\rm B}}$.
This additional factor modifies Eq.~\ref{eq:integrated_lum}, introducing a different dependence of \llat\ on \ee\ and \eb.
The dependence from \ee\ becomes weaker, while the one from \eb\ (that was negligible) becomes stronger: $L_{LAT}\propto \epsilon_{\rm e}^{p-3/2} \epsilon_{\rm B}^{p/4}\simeq\sqrt{\epsilon_{\rm e} \epsilon_{\rm B}}$.
However, for high-energy photons the KN limit can be relevant and should be taken into account. If this is the case, $Y$ is no longer constant but depends on the electron Lorentz factor $\gamma_e$.

Following \cite{nakar09}, we introduce the quantity $\widehat\gamma_m=m_ec^2\Gamma/h\nu_m$:
photons with energy larger than $h\nu_m$ cannot be efficiently upscattered by electrons with $\gamma_e > \widehat\gamma_m$, because they are above the KN limit. 
In fast cooling, the importance of the KN effects is determined by the ratio $\widehat\gamma_{m}/\gamma_m$.
When $\widehat\gamma_{m}<\gamma_m$ the synchrotron spectrum is in the strong KN regime. This condition is verified up to $t=470\epsilon_{e,-1}^2\xi_{e,-1}^{-2}\epsilon_{B,-2}^{1/3}E_{K,54}^{1/3}\,$s. 
In this regime the shape of the spectrum depends on the relation between $\widehat\gamma_{m}/\gamma_m$ and \ee/\eb\ but, in all cases, the part of the spectrum above $max(\nu_m,\nu_m\frac{\epsilon_e}{\epsilon_B}\frac{\widehat\gamma_m}{\gamma_m})$ is strongly affected by KN, and SSC losses do not significantly modify the synchrotron spectrum.
The LAT energy range ($0.1-10\,$GeV) is always in this regime, since it is above $h\nu_m$ (which is not modified by SSC and is still given by Eq.~\ref{eq:Em}) and it is above $h\nu_m\frac{\epsilon_e}{\epsilon_B}\frac{\widehat\gamma_m}{\gamma_m}$ for $\epsilon_B > 1.7\times10^{-6}\xi_{e,-1}$. 

Following \cite{nakar09} we have also estimated the contribution of the SSC spectral component to the flux in the LAT energy range. 
Since \eb\ is the most uncertain parameter, we fix the value of all the other parameters to the typical values used in the previous equations and vary \eb\ in the range $10^{-2} -10^{-5}$. We find that the SSC component never dominates the LAT emission over the synchrotron one. Due to a reduction of KN effects, the importance of the SSC component in the LAT range increases with time and for smaller \eb. However, at small enough \eb ($\epsilon_{\rm B}\lesssim10^{-5}$) the transition to the slow cooling regime occurs at times as early as 200 s and reduces the importance of IC losses.
Similar conclusions have been reached by \cite{wang10}.
Even if the SSC emission never dominates over the synchrotron in the LAT energy range, we found that, depending on the model parameters, the SSC photon flux can be high enough to explain the detection of a few photons at energy in excess of $\sim10\,$GeV at late time \citep{wang13,tang14}.

\section{Discussion}\label{sect:discussion}
In this section we discuss our findings on the distributions of the parameters \ee\ and \etag.

\subsection{Clustering of X-ray and optical light curves}\label{sec:otherworks}
\begin{figure}
\includegraphics[scale=0.45]{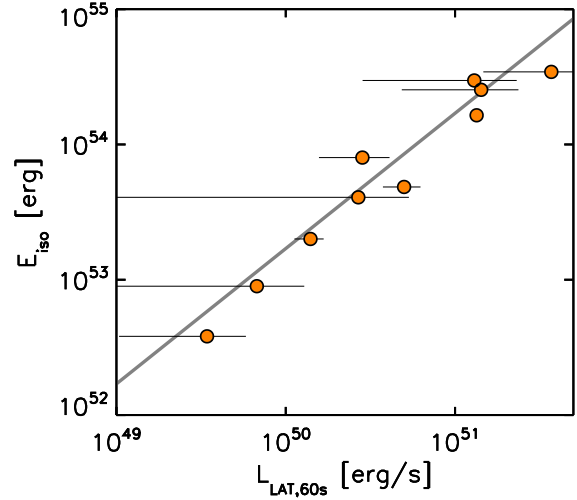}
\caption{Correlation between \eiso\ and the LAT luminosity $L_{LAT}(t)$ estimated at $t=60\,$s. The solid line has slope 1 and it is shown for reference.}
\label{fig:correla_60sec}
\end{figure}
This paper reports on the existence of a correlation between the LAT luminosity and the prompt energetics, and suggests that (under the assumption that LAT radiation is synchrotron emission from ambient electrons accelerated in the external shock) such a relation can be used to infer the width of the distributions of \ee\ and \etag.
Although this is the first time that the relation between LAT luminosity and $E_{\rm prompt}$ is used to infer the properties of these parameters, similar analyses have already been performed using afterglow data at different frequencies and at later times.
In these studies, the narrowness of the ratio $L_{\rm aft}/E_{\rm prompt}$ (where $L_{\rm aft}$ is the afterglow luminosity) is not usually represented in terms of a clustering of the \eiso-normalised light curves but, equivalently, as a linear correlation between \eiso\ and the afterglow luminosity estimated at some fixed time (often around 10-24 hours).
Several examples of these kinds of studies can be found in literature.
\cite{berger07} found a correlation between the X-ray luminosities at one day and \eiso\ in a sample of 16 short GRBs and concluded that  this finding implies narrow distributions for \etag\ and \ee. An updated version of this correlation and the comparison with long bursts, both in X-ray and optical bands can be found in \cite{berger13}.
Similar conclusions were reached by \citet{kaneko07} on a sample of 27 regular long GRBs \citep{nousek06} and 4 GRBs associated with SNe.
\citet{davanzo12} considered the BAT6 sample, a sub-sample of {\it Swift}/BAT GRBs almost complete in redshift  \citep{salvaterra12}, and studied how the $L_X(t)-E_{\rm prompt}$ correlation changes over time, from 5 minutes to 1 day.
They found that, even if a significant correlation is always present, its dispersion increases with time. The same conclusions have been reached by \cite{margutti13}, who studied the $L_X(t)-E_{\rm prompt}$ correlation for large samples of both long and short GRBs at $t=10\,\rm min$ and $t=11\,\rm hr$. 
The observed weakening of the correlation at late time is not surprising.
As pointed out by \cite{kumar00}, the dispersion is expected to increase due to the contribution of $\sigma_p$: since $p$ enters not only the normalization but also the slope of the light curves (see Eq.~\ref{eq:mocro_lum}), its contribution to the dispersion of the ratio $L_{aft}/E_{\rm prompt}$ increases with time (see \citealt{kumar00} for a detailed estimate of this effect and its dependence on the time and frequency of observations).

Another source of scattering is the possible presence, in the considered sample, of GRBs for which the observed frequency at the time of observations is below $\nu_c$, where the flux depends also on the density.
Depending on the parameters, $\nu_c$ is expected to eventually cross the X-ray energy band at different times for different bursts.
\cite{berger07}, for example, found that a small fraction of short bursts does not follow the strong $L_X-E_{\rm prompt}$ correlation defined by the majority of the bursts in its sample and concluded that these bursts have a low circumstellar density, leading to $\nu_c>\nu_X$ at 1 day (see also \citealt{nakar07}).

In Fig.~\ref{fig:correla_60sec} we show the correlation between \eiso\ and the LAT luminosity at 60 seconds, $L_{LAT,60}$. 
The advantages of using high-energy data are many.
LAT data are available at early time and, as shown by \citet{kumar00}, the contribution of $\sigma_p$ to the width of \lovere\ is smaller at early time and quickly increases at later time.
Also, it is very likely that the LAT energy range lies above the typical synchrotron frequencies, avoiding contamination from observations at frequencies where the luminosity is not a good proxy for $E_{\rm K}$.
Caveats to the use of high-energy data are discussed in Section \ref{sec:biases}

\subsection{Correlation between $\epsilon_e$ and $\eta_\gamma$}\label{sec:correlation}
The statement that the correlation between afterglow luminosity and prompt energetics implies a narrow distribution of both \ee\ and \etag\ is based on the assumption that these two parameters are not correlated. In this section we relax the assumption of independence.
In this case, it is still possible to reproduce the clustering provided that the product $\epsilon_e(1-\eta_\gamma)/\eta_\gamma$ remains constant.
This means that the two parameters are correlated, and track each other (i.e., when \etag\ is larger, then also \ee\ should be larger).
In this case they are not required to have narrow distributions and they can vary across a wide range.
The efficiency \etag\ describes the conversion of jet energy (in a kinetic or magnetic form) into observed radiation and is the product of several factors.
Limiting our discussion to the internal shock scenario, the flow kinetic energy is dissipated into internal energy with efficiency $\eta_{diss}$, then electrons are accelerated by the shock with efficiency $\epsilon_{e,\gamma}$ and they radiate via synchrotron emission with efficiency $\eta_{rad}\sim1$ (fast cooling regime). The overall efficiency is then given by $\eta_\gamma=\eta_{diss}\epsilon_{e,\gamma}\eta_{rad}$. In this internal-external shock scenario, particles radiating the prompt and afterglow emission are accelerated in both cases via collisionless shocks, and this can explain why the two efficiencies track each other. 
However, the efficiency of the two processes can be very different and not necessarily related to each other, since internal shocks are mildly relativistic and may be magnetised, and their physics could be very different form the physics of the external shocks.
Due to our poor knowledge of the mechanism at work in the prompt phase, it is difficult to argue in favour or against a correlation between \ee\ and \etag. 
Moreover, recent particle-in-cell (PIC) simulations showed that for ultra-relativistic ($\Gamma_0>10$) weakly magnetised ($\sigma<10^{-5}$) shocks, as expected in external shocks in GRBs, the acceleration efficiency does not show any dependence on the flow energy (or on the external density or on the magnetisation as well) and $\epsilon_e$ is clustered around a typical value $\sim10\%$ \citep{sironi13}. This is in agreement with our results and suggests that it is reasonable to assume that \ee\ has a narrow distribution, and is not correlated to other quantities. This implies that also \etag\ should be narrowly distributed. 
Summarising, both data modelling and numerical simulations support the existence of a typical value for the acceleration efficiency at external shocks, favouring the scenario in which the clustering can be explained only invoking the existence of a typical value also for the prompt efficiency.

\subsection{Presence of selection biases}\label{sec:biases}
Several studies of the $L_{\rm aft}-E_{\rm prompt}$ relation at X-ray and optical frequencies have reached conclusions (about the narrowness of \ee\ and \etag) similar to the ones derived in this paper.
However, the widths derived from LAT data are smaller than the ones derived from analysis performed at different frequencies. 
As anticipated, this is partially due to the role of $\sigma_p$, whose contribution to the dispersion of the $L_{\rm aft}-E_{\rm prompt}$ relation increases with time. Also, the small number of the events considered here might of course contribute to underestimate the dispersion of the $L_{LAT}-E_{\rm prompt}$ relation.
Besides, the sample we are considering is constituted by GRBs detected by LAT and with measured redshift. Both these requirements introduce a selection effect that favours powerful bursts. If compared with their parent population, the bursts in our sample lie in the intermediate/high-values part of the \eiso\ and luminosity distributions, i.e. they are not representative of the whole GRB population. 
This selection bias can affect the results. 
In particular, it is possible that the very narrow distributions derived here are due to the fact that we are sampling only a part of the whole distributions of \ee\ and \etag. 
However, this last statement is true only if these parameters are correlated with the GRB energetics/luminosity.

As discussed in the previous section, the predicted value of \ee\ from PIC simulations is robust and independent on other parameters, and then characterised by a small dispersion. No correlation with other quantities is found provided that the bulk Lorentz factor is larger than $\sim 10$. Then, very high-energy bursts should not show any difference from the weakest ones in terms of \ee\ and should not introduce any bias in the constraints derived for this parameter.

A correlation between \eiso\ and \etag\ is instead very likely. 
Even if the nature of the mechanism that converts the jet energy into prompt radiation is uncertain, bursts characterised by high energy outputs are those bursts for which the mechanism for energy extraction has been particularly efficient.
This means that the sample we are considering is a sub-sample of bursts with high values of \etag, which may not be representative of the whole GRB population and this can explain why we derived a very narrow distribution for this parameter.

\section{Conclusions}\label{sect:conclusions}
Strong correlations between \eiso\ and the afterglow luminosity (measured at a fixed time $t$) have been reported by several authors, both for X-ray and optical luminosities, for $t$ that varies from a few minutes to several hours. 
These correlations have been used to argue that the value of \ee\ and \etag\ must be narrowly distributed, since only in this case the afterglow luminosity can be a good proxy for the energy released during the prompt phase. 
This conclusion is derived from interpreting the emission as synchrotron radiation from external shocks. 
This analysis is usually performed using X-ray observations at late time, when the X-ray band possibly falls above the typical synchrotron frequencies, where the luminosity is independent from the density and only weakly dependent on \eb. 
The optical band instead falls more likely below the cooling frequency, where the luminosity depends also on these parameters.

In this paper we report on a similar correlation found between \eiso\ and LAT luminosities. This correlation is strong no matter the time at which the LAT luminosity is estimated. For this reason, this result can be represented as a clustering of the \eiso-normalised LAT light curves, i.e., a decrease in the dispersion between the light curves of different bursts, once they are re-normalised using \eiso. 
The relevance of this result is twofold.
On the one hand, this finding (first reported by \citealt{ghisellini10} with a sample of four GRBs and then confirmed in this work with a sample of ten GRBs) gives strong support to the interpretation of the long-lasting GeV emission as synchrotron radiation produced at the external shock. On the other hand, the study of the small dispersion of the \eiso-normalised LAT light curves allows us to derive strong constraints on the distributions of \ee\ and \etag. 

In this paper we focused first on the possibility of modelling LAT light curves with the standard synchrotron/external-shock model. We derived that:
\begin{itemize}
\item the estimate of the synchrotron flux in the LAT energy range is not affected by SSC cooling, since the process for up-scattering of LAT photons proceeds in KN regime and is strongly suppressed;
\item the synchrotron luminosity predicted in the LAT range (eq.~\ref{eq:integrated_lum}) is consistent with the measured luminosity (Fig.~\ref{fig:clustering});
\item the observed flux decay rate and the spectral shape are consistent with predictions;
\item using the parameters for which LAT emission can be modelled as synchrotron radiation, the predicted SSC component does not dominate the LAT flux over the synchrotron;
\item the validity of the previous statements has been discussed for different values of \eb, from the fiducial one ($\epsilon_{\rm B}\sim10^{-2}$) to the smaller ones recently suggested by broad band afterglow modelling.
\end{itemize}

Since we showed that observations are consistent with synchrotron emission, we therefore assumed that this mechanism is responsible for the high-energy radiation and we derived that:
\begin{itemize}
\item \ee\ and \etag\ have narrow distributions;
\item the maximum value for $\sigma_{Log\,\epsilon}$ is 0.19 (Fig.~\ref{fig:dispersion});
\item the maximum value for $\sigma_{Log\,\eta}$ is 0.23 if $\langle\eta_\gamma\rangle\ll1$, but it is sensitively smaller for higher values of $\langle\eta_\gamma\rangle$ (Fig.~\ref{fig:dispersion}).
\end{itemize}

The unprecedented energy coverage and sensitivity provided by LAT showed that the spectral and temporal properties of the emission from GRBs 
are characterised by several recurrent features common to most GRBs.
The phenomenological result described in this paper, i.e. the strong and universal relation between the LAT luminosity during the power-law decay phase and the prompt energetics, should be considered as one additional property characterising the high-energy radiation in GRBs, at least in those cases in which a long-lasting emission is detected.
Any theoretical model aimed at interpreting the origin of the temporally extended high-energy emission must be able to explain all these observations. 
In this paper we have focused our discussion on the scenario in which the extended emission is dominated by synchrotron radiation from external shocks and we have demonstrated that all these features, included the one presented in this paper, can be easily explained.

\section*{Acknowledgements}
LN was supported by a Marie Curie Intra-European Fellowship of the European Community's 7th Framework Programme (PIEF-GA-2013-627715).
LN and RBD were supported by an ERC advanced grant (GRB) and by the I-CORE Program of the PBC and the ISF (grant 1829/12).

\bibliography{biblio.bib}

\begin{thebibliography}{}

\bibitem[\protect\citeauthoryear{{Ackermann}, {Ajello}, {Asano}, {Atwood},
  {Axelsson}, {Baldini}, {Ballet}, {Barbiellini} \& {Baring}}{{Ackermann}
  et~al.}{2014}]{130427LAT}
{Ackermann} M.,  {Ajello} M.,  {Asano} K.,  {Atwood} W.~B.,  {Axelsson} M.,
  {Baldini} L.,  {Ballet} J.,  {Barbiellini} G.,    {Baring} M.~G. e.~a.,
  2014, Science, 343, 42

\bibitem[\protect\citeauthoryear{{Ackermann}, {Ajello}, {Asano}, {Axelsson},
  {Baldini}, {Ballet}, {Barbiellini}, {Bastieri}, {Bechtol}, {Bellazzini} \&
  {Bhat}}{{Ackermann} et~al.}{2013}]{latcatalog}
{Ackermann} M.,  {Ajello} M.,  {Asano} K.,  {Axelsson} M.,  {Baldini} L.,
  {Ballet} J.,  {Barbiellini} G.,  {Bastieri} D.,  {Bechtol} K.,  {Bellazzini}
  R.,    {Bhat} e.~a.,  2013, \apjs, 209, 11

\bibitem[\protect\citeauthoryear{{Atwood}, {Abdo}, {Ackermann}, {Althouse},
  {Anderson}, {Axelsson}, {Baldini}, {Ballet}, {Band}, {Barbiellini} \& et
  al.}{{Atwood} et~al.}{2009}]{atwood09}
{Atwood} W.~B.,  {Abdo} A.~A.,  {Ackermann} M.,  {Althouse} W.,  {Anderson} B.,
   {Axelsson} M.,  {Baldini} L.,  {Ballet} J.,  {Band} D.~L.,  {Barbiellini}
  G.,    et al. 2009, \apj, 697, 1071

\bibitem[\protect\citeauthoryear{{Barniol Duran}}{{Barniol
  Duran}}{2013}]{barniolduran13}
{Barniol Duran} R.,  2013, ArXiv:1311.1216

\bibitem[\protect\citeauthoryear{{Beloborodov}, {Hascoet} \&
  {Vurm}}{{Beloborodov} et~al.}{2013}]{beloborodov13}
{Beloborodov} A.~M.,  {Hascoet} R.,    {Vurm} I.,  2013, ArXiv:1307.2663

\bibitem[\protect\citeauthoryear{{Berger}}{{Berger}}{2007}]{berger07}
{Berger} E.,  2007, \apj, 670, 1254

\bibitem[\protect\citeauthoryear{{Berger}}{{Berger}}{2013}]{berger13}
{Berger} E.,  2013, ArXiv e-prints

\bibitem[\protect\citeauthoryear{{D'Avanzo}, {Salvaterra}, {Sbarufatti},
  {Nava}, {Melandri}, {Bernardini}, {Campana}, {Covino}, {Fugazza},
  {Ghirlanda}, {Ghisellini}, {Parola}, {Perri}, {Vergani} \&
  {Tagliaferri}}{{D'Avanzo} et~al.}{2012}]{davanzo12}
{D'Avanzo} P.,  {Salvaterra} R.,  {Sbarufatti} B.,  {Nava} L.,  {Melandri} A.,
  {Bernardini} M.~G.,  {Campana} S.,  {Covino} S.,  {Fugazza} D.,  {Ghirlanda}
  G.,  {Ghisellini} G.,  {Parola} V.~L.,  {Perri} M.,  {Vergani} S.~D.,
  {Tagliaferri} G.,  2012, \mnras, 425, 506

\bibitem[\protect\citeauthoryear{{De Pasquale}, {Schady}, {Kuin}, {Page},
  {Curran}, {Zane}, {Oates}, {Holland}, {Breeveld}, {Hoversten}, {Chincarini},
  {Grupe}, {Abdo}, {Ackermann}, {Ajello}, {Axelsson}, {Baldini} \&
  {Ballet}}{{De Pasquale} et~al.}{2010}]{depasquale10}
{De Pasquale} M.,  {Schady} P.,  {Kuin} N.~P.~M.,  {Page} M.~J.,  {Curran}
  P.~A.,  {Zane} S.,  {Oates} S.~R.,  {Holland} S.~T.,  {Breeveld} A.~A.,
  {Hoversten} E.~A.,  {Chincarini} G.,  {Grupe} D.,  {Abdo} A.~A.,  {Ackermann}
  M.,  {Ajello} M.,  {Axelsson} M.,  {Baldini} L.,    {Ballet} J. e.~a.,  2010,
  \apjl, 709, L146

\bibitem[\protect\citeauthoryear{{Gao}, {Mao}, {Xu} \& {Fan}}{{Gao}
  et~al.}{2009}]{gao09}
{Gao} W.-H.,  {Mao} J.,  {Xu} D.,    {Fan} Y.-Z.,  2009, \apjl, 706, L33

\bibitem[\protect\citeauthoryear{{Ghirlanda}, {Ghisellini} \&
  {Nava}}{{Ghirlanda} et~al.}{2010}]{ghirlanda10}
{Ghirlanda} G.,  {Ghisellini} G.,    {Nava} L.,  2010, \aap, 510, L7

\bibitem[\protect\citeauthoryear{{Ghisellini}, {Ghirlanda}, {Nava} \&
  {Celotti}}{{Ghisellini} et~al.}{2010}]{ghisellini10}
{Ghisellini} G.,  {Ghirlanda} G.,  {Nava} L.,    {Celotti} A.,  2010, \mnras,
  403, 926

\bibitem[\protect\citeauthoryear{{Granot} \& {Sari}}{{Granot} \&
  {Sari}}{2002}]{granot02}
{Granot} J.,  {Sari} R.,  2002, \apj, 568, 820

\bibitem[\protect\citeauthoryear{{Kaneko}, {Ramirez-Ruiz}, {Granot},
  {Kouveliotou}, {Woosley}, {Patel}, {Rol}, {in 't Zand}, {van der Horst},
  {Wijers} \& {Strom}}{{Kaneko} et~al.}{2007}]{kaneko07}
{Kaneko} Y.,  {Ramirez-Ruiz} E.,  {Granot} J.,  {Kouveliotou} C.,  {Woosley}
  S.~E.,  {Patel} S.~K.,  {Rol} E.,  {in 't Zand} J.~J.~M.,  {van der Horst}
  A.~J.,  {Wijers} R.~A.~M.~J.,    {Strom} R.,  2007, \apj, 654, 385

\bibitem[\protect\citeauthoryear{{Kumar}}{{Kumar}}{2000}]{kumar00}
{Kumar} P.,  2000, \apjl, 538, L125

\bibitem[\protect\citeauthoryear{{Kumar} \& {Barniol Duran}}{{Kumar} \&
  {Barniol Duran}}{2009}]{kumar09}
{Kumar} P.,  {Barniol Duran} R.,  2009, \mnras, 400, L75

\bibitem[\protect\citeauthoryear{{Kumar} \& {Barniol Duran}}{{Kumar} \&
  {Barniol Duran}}{2010}]{kumar10}
{Kumar} P.,  {Barniol Duran} R.,  2010, \mnras, 409, 226

\bibitem[\protect\citeauthoryear{{Lemoine}, {Li} \& {Wang}}{{Lemoine}
  et~al.}{2013}]{lemoine13}
{Lemoine} M.,  {Li} Z.,    {Wang} X.-Y.,  2013, \mnras, 435, 3009

\bibitem[\protect\citeauthoryear{{Margutti}, {Zaninoni}, {Bernardini},
  {Chincarini}, {Pasotti}, {Guidorzi}, {Angelini}, {Burrows}, {Capalbi},
  {Evans}, {Gehrels}, {Kennea}, {Mangano} \& {Moretti}}{{Margutti}
  et~al.}{2013}]{margutti13}
{Margutti} R.,  {Zaninoni} E.,  {Bernardini} M.~G.,  {Chincarini} G.,
  {Pasotti} F.,  {Guidorzi} C.,  {Angelini} L.,  {Burrows} D.~N.,  {Capalbi}
  M.,  {Evans} P.~A.,  {Gehrels} N.,  {Kennea} J.,  {Mangano} V.,    {Moretti}
  A. e.~a.,  2013, \mnras, 428, 729

\bibitem[\protect\citeauthoryear{{Maxham}, {Zhang} \& {Zhang}}{{Maxham}
  et~al.}{2011}]{maxham11}
{Maxham} A.,  {Zhang} B.-B.,    {Zhang} B.,  2011, \mnras, 415, 77

\bibitem[\protect\citeauthoryear{{Meegan}, {Lichti}, {Bhat}, {Bissaldi},
  {Briggs}, {Connaughton}, {Diehl}, {Fishman}, {Greiner}, {Hoover} \& {van der
  Horst}}{{Meegan} et~al.}{2009}]{meegan09}
{Meegan} C.,  {Lichti} G.,  {Bhat} P.~N.,  {Bissaldi} E.,  {Briggs} M.~S.,
  {Connaughton} V.,  {Diehl} R.,  {Fishman} G.,  {Greiner} J.,  {Hoover} A.~S.,
     {van der Horst} A.~J. e.~a.,  2009, \apj, 702, 791

\bibitem[\protect\citeauthoryear{{Nakar}}{{Nakar}}{2007}]{nakar07}
{Nakar} E.,  2007, \physrep, 442, 166

\bibitem[\protect\citeauthoryear{{Nakar}, {Ando} \& {Sari}}{{Nakar}
  et~al.}{2009}]{nakar09}
{Nakar} E.,  {Ando} S.,    {Sari} R.,  2009, \apj, 703, 675

\bibitem[\protect\citeauthoryear{{Nava}, {Vianello}, {Omodei}, {Ghisellini},
  {Ghirlanda}, {Celotti}, {Longo} \& {Desiante}}{{Nava} et~al.}{2013}]{nava13}
{Nava} L.,  {Vianello} G.,  {Omodei} N.,  {Ghisellini} G.,  {Ghirlanda} G.,
  {Celotti} A.,  {Longo} F.,    {Desiante} R.,  2013, ArXiv:1308.5442

\bibitem[\protect\citeauthoryear{{Nousek}, {Kouveliotou}, {Grupe}, {Page},
  {Granot}, {Ramirez-Ruiz}, {Patel}, {Burrows}, {Mangano}, {Barthelmy},
  {Beardmore}, {Campana}, {Capalbi}, {Chincarini}, {Cusumano}, {Falcone} \&
  {Gehrels}}{{Nousek} et~al.}{2006}]{nousek06}
{Nousek} J.~A.,  {Kouveliotou} C.,  {Grupe} D.,  {Page} K.~L.,  {Granot} J.,
  {Ramirez-Ruiz} E.,  {Patel} S.~K.,  {Burrows} D.~N.,  {Mangano} V.,
  {Barthelmy} S.,  {Beardmore} A.~P.,  {Campana} S.,  {Capalbi} M.,
  {Chincarini} G.,  {Cusumano} G.,  {Falcone} A.~D.,    {Gehrels} 2006, \apj,
  642, 389

\bibitem[\protect\citeauthoryear{{Panaitescu} \& {Kumar}}{{Panaitescu} \&
  {Kumar}}{2000}]{panaitescu00}
{Panaitescu} A.,  {Kumar} P.,  2000, \apj, 543, 66

\bibitem[\protect\citeauthoryear{{Panaitescu} \& {Kumar}}{{Panaitescu} \&
  {Kumar}}{2001}]{panaitescu01}
{Panaitescu} A.,  {Kumar} P.,  2001, \apjl, 560, L49

\bibitem[\protect\citeauthoryear{{Piran} \& {Nakar}}{{Piran} \&
  {Nakar}}{2010}]{piran10}
{Piran} T.,  {Nakar} E.,  2010, \apjl, 718, L63

\bibitem[\protect\citeauthoryear{{Salvaterra}, {Campana}, {Vergani}, {Covino},
  {D'Avanzo}, {Fugazza}, {Ghirlanda}, {Ghisellini}, {Melandri}, {Nava},
  {Sbarufatti}, {Flores}, {Piranomonte} \& {Tagliaferri}}{{Salvaterra}
  et~al.}{2012}]{salvaterra12}
{Salvaterra} R.,  {Campana} S.,  {Vergani} S.~D.,  {Covino} S.,  {D'Avanzo} P.,
   {Fugazza} D.,  {Ghirlanda} G.,  {Ghisellini} G.,  {Melandri} A.,  {Nava} L.,
   {Sbarufatti} B.,  {Flores} H.,  {Piranomonte} S.,    {Tagliaferri} G.,
  2012, \apj, 749, 68

\bibitem[\protect\citeauthoryear{{Santana}, {Barniol Duran} \&
  {Kumar}}{{Santana} et~al.}{2013}]{santana14}
{Santana} R.,  {Barniol Duran} R.,    {Kumar} P.,  2013, ArXiv e-prints

\bibitem[\protect\citeauthoryear{{Sari} \& {Esin}}{{Sari} \&
  {Esin}}{2001}]{sari01}
{Sari} R.,  {Esin} A.~A.,  2001, \apj, 548, 787

\bibitem[\protect\citeauthoryear{{Sari}, {Piran} \& {Narayan}}{{Sari}
  et~al.}{1998}]{sari98}
{Sari} R.,  {Piran} T.,    {Narayan} R.,  1998, \apjl, 497, L17

\bibitem[\protect\citeauthoryear{{Sironi}, {Spitkovsky} \& {Arons}}{{Sironi}
  et~al.}{2013}]{sironi13}
{Sironi} L.,  {Spitkovsky} A.,    {Arons} J.,  2013, \apj, 771, 54

\bibitem[\protect\citeauthoryear{{Tang}, {Tam} \& {Wang}}{{Tang}
  et~al.}{2014}]{tang14}
{Tang} Q.-W.,  {Tam} P.-H.~T.,    {Wang} X.-Y.,  2014, \apj, 788, 156

\bibitem[\protect\citeauthoryear{{Vurm}, {Hascoet} \& {Beloborodov}}{{Vurm}
  et~al.}{2014}]{vurm14}
{Vurm} I.,  {Hascoet} R.,    {Beloborodov} A.~M.,  2014, ArXiv:1402.2595

\bibitem[\protect\citeauthoryear{{Wang}, {He}, {Li}, {Wu} \& {Dai}}{{Wang}
  et~al.}{2010}]{wang10}
{Wang} X.-Y.,  {He} H.-N.,  {Li} Z.,  {Wu} X.-F.,    {Dai} Z.-G.,  2010, \apj,
  712, 1232

\bibitem[\protect\citeauthoryear{{Wang}, {Liu} \& {Lemoine}}{{Wang}
  et~al.}{2013}]{wang13}
{Wang} X.-Y.,  {Liu} R.-Y.,    {Lemoine} M.,  2013, \apjl, 771, L33

\bibitem[\protect\citeauthoryear{{Wijers} \& {Galama}}{{Wijers} \&
  {Galama}}{1999}]{wijers99}
{Wijers} R.~A.~M.~J.,  {Galama} T.~J.,  1999, \apj, 523, 177

\end{thebibliography}

\label{lastpage}

\end{document}